\documentclass[aps,prb,twocolumn,groupedaddress,showpacs,superscriptaddress,amssymb,amsmath]{revtex4-1}
\usepackage{graphicx}
\usepackage{tabularx}
\usepackage{color}
\usepackage{amsmath}
\usepackage{mathtools}
\usepackage{comment}
\newcommand{\be}{\begin{equation}}
\newcommand{\ee}{\end{equation}}
\newcommand{\bea}{\begin{eqnarray}}
\newcommand{\eea}{\end{eqnarray}}
\usepackage{dcolumn}
\usepackage{hyperref}
\usepackage{bm}
\usepackage{epsf}
\usepackage{subcaption}
\usepackage{float}
\usepackage{epstopdf}%
\usepackage{amsfonts}
\usepackage{braket}
\usepackage{cancel}

\DeclareMathOperator{\trace}{Tr}

\begin{document}


\title{Cooling condition for multilevel quantum absorption refrigerators}

\author{Hava Meira Friedman}
\address{Chemical Physics Theory Group, Department of Chemistry, University of Toronto, 80 Saint George St., Toronto, Ontario, Canada M5S 3H6}
\author{Dvira Segal}
\address{Chemical Physics Theory Group, Department of Chemistry, University of Toronto, 80 Saint George St., Toronto, Ontario, Canada M5S 3H6}
\address{Department of Physics, 60 Saint George St., University of Toronto, Toronto, Ontario, Canada M5S 1A7} 
\date{\today}

\begin{abstract}
Models for quantum absorption refrigerators serve as test beds for exploring concepts and 
developing methods in quantum thermodynamics. 
Here, we depart from the minimal, ideal design and consider a 
generic multilevel model for a quantum absorption refrigerator, which potentially suffers from lossy processes.
Based on a full-counting statistics approach, we derive a formal cooling condition for the refrigerator,
 which can be feasibly evaluated analytically and numerically.
We exemplify our approach on a three-level model for a quantum absorption refrigerator that suffers
from different forms of non-ideality (heat leakage, competition between different cooling pathways), 
and examine the cooling current with different designs.
This study assists in identifying the cooling window of imperfect thermal machines.
\end{abstract}

\maketitle

\section{Introduction}
\label{sec-intro}

Recent years have seen an explosion in research aiming to tie the fields of quantum mechanics 
and classical thermodynamics, with a focus on quantum thermal machines \cite{AndersRev,LevyRev,bookKos}. 
Of particular interest is the quantum absorption refrigerator (QAR) 
\cite{Scovil-Schluz-Dubois,Kosloff2001,Levy2012,Skrzypczyk2010,Correa2013,Correa2014Nature,AlonsoNJP17,Dvira2018,Mitchison-rev}, 
a machine, analogous to the classical counterpart, which employs a heat source rather than an 
external power input to achieve refrigeration of a target component. 
Here, the ``working fluid" that converts energy to refrigerate is quantum mechanical in nature. 
QARs are particularly useful in nanotechnological applications because they utilize waste heat and
 operate autonomously without driving.  
Potential applications include quantum state preparation, quantum computing \cite{natphys2015}, 
and biological function in proteins \cite{Popescu2013}.  

Understanding the working principles of QARs helps consolidate the connection between 
the theories of quantum mechanics and thermodynamics, 
just as the steam engine was instrumental to the development of classical thermodynamics  \cite{Carnot}. 
Numerous studies examine the role of quantum effects in nanoscale absorption refrigerators, 
including quantum coherences \cite{NJP2015,Brunner,Michael2018,Noise2018,Pet,Zambrini,QCrev}, 
quantum information resources \cite{QI}, 
strong system-bath coupling 
\cite{Tanimura16,Strasberg16,NJP2018, Anqi,Tanimura19,Yun19}, strong internal couplings \cite{Seah,Du}
and bath engineering \cite{Correa2014Nature}. 
Experimental realizations for QARs were recently proposed \cite{BrunnerPRB2016,Plenio2016,Pekola}, 
with the first implementation utilizing trapped ions described in Ref. \cite{NatComm2019}. 
The behavior of specific model systems for QARs were explored in many studies, including
\cite{Correa2013,Correa2014,Correa2015PRE,BrunnerPRE2015,WangPRE2015,Barra2017,Scarani2017,NJP2018}.


Different mechanisms are responsible for irreversibility in QARs, impacting performance bounds. 
This includes heat leaks: the parasitic coupling of heat baths to the cooling and driving transitions, 
internal dissipation, which corresponds to the competition between different cooling pathways,
and delocalized dissipation. 
The effect of such lossy mechanisms on the performance of QARs was discussed in specific model systems, see e.g. Refs. 
\cite{Correa2013,Correa2014}.
%
A graph theory analysis of multistate QARs (with their dynamics described by classical rate equations)
was recently presented in Ref. \cite{AlonsoNJP17}. This treatment allows decomposition of 
the cooling current in terms of the circuits of the machine, bringing in strategies to enhance the cooling performance.
Most notably, it was demonstrated in Ref. \cite{AlonsoNJP17} that the performance of incoherent multilevel QAR was 
smaller than or equal to the performance of the best-performing circuit component.
A graph theory treatment expresses the cooling current as a decomposition of different cycles.
Nevertheless, a framework for efficiently 
resolving the cooling window, cooling current and associated noise of a  QAR device {\it directly from the generator 
of the dynamics} is highly desirable. 
Particularly, given efforts to realize QARs, 
a viable  analysis of the performance of imperfect devices suffering lossy processes is crucial.

In this work, we show that a cooling condition for QARs can be defined quite generally 
and analytically for a multilevel system coupled to multiple heat baths, 
where one of the baths (labeled $C$) is refrigerated by the setup. 
The flexibility of our analytical results allows us to demonstrate 
that a three-level QAR \cite{Scovil-Schluz-Dubois,Kosloff2001,Skrzypczyk2010} can operate 
even in the presence of heat leaks and internal dissipation to all baths in the setup. 
More broadly, our formalism allows us to efficiently calculate, analytically and numerically, heat exchange  
in devices with multiplexed couplings to thermal reservoirs using linear algebra operations 
applied directly on the Liouvillian of the dynamics.
These results are achieved using an open quantum system full-counting statistics approach.

The paper is organized as follows.
In Section \ref{sec:cooling}, we introduce the non-ideal multilevel refrigerator and derive
expressions for its cooling current and the cooling condition 
using a truncated cumulant generating function. 
Theoretical calculations are exemplified with simulations of a three-level QAR in Sec. \ref{sec:3LQAR}.
Further examples and details are delegated to Appendixes A and B.
We conclude and discuss future directions in Section \ref{sec:summ}.

\section{Analytic Results}
\label{sec:cooling}

\subsection{Setup and equations of motion}

\begin{figure}[H]
\includegraphics[width=6.5cm]{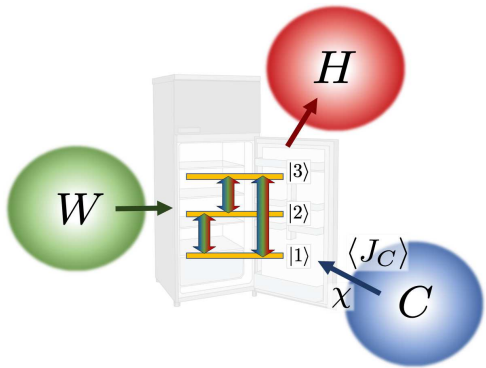}
\vspace{-2mm}
\hspace{10mm}
\caption{
A scheme of the model refrigerator. The three-level system acts as a working material.
Transitions between levels are induced by the heat baths.
Unlike the optimal design \cite{Levy2012,Correa2014Nature}, 
several reservoirs may simultaneously couple to the same transition.}
\label{fig:scheme}
\end{figure}

Our setup includes an $N$-level, nondegenerate system, 
which is coupled to multiple heat baths (enumerated by $\mu$) 
maintained at different temperatures.
These baths are responsible for inducing transitions (excitations and relaxations) between levels, $N\geq 2$. 
We specifically identify the coldest heat bath (denoted by $C$); our goal is to 
extract heat from this reservoir, assisted by other baths, and dump the heat into other heat reservoirs.
Note that only thermal energy (heat) is exchanged here between system and baths.
A three-level example is displayed in Figure (\ref{fig:scheme}). 

We are interested in the steady state behavior of this nonequilibrium system. 
To this end, we make several standard assumptions: 
weak system-bath coupling, factorized system-bath initial state, 
Markovian dynamics, and the secular approximation  (decoupling population and coherences). 
The latter approximation is valid when  energy levels of the system
are nondegenerate and the temperature is high. 
Under these assumptions, the dynamics of the $N$-level system can be organized as a 
Markovian quantum master equation (MQME),
an equation of motion for the system's population, $p$ \cite{OQSbook},
\bea
\dot p(t) = {\cal L} \, p(t).
\label{eq:qme}
\eea
Here, ${\cal L}$ is a Liouvillian matrix of size $N$ by $N$, which is built from bath-induced rate constants 
that lead to changes in the system's state populations. 
Under the weak-coupling approximation, in additive interaction models 
${\cal L}$ is given by a sum of the Liouvillians (transitions rates) for each bath, $\sum_\mu {\cal L}_\mu$. 

Given this out-of-equilibrium, $N$-level, multi-bath model, our objective is to calculate the 
heat current extracted from the cold bath. 
We approach this challenge using a full-counting statistics (FCS) formalism 
\cite{EspositoReview,hanggi-review,bijay-wang-review}.
This tool provides the cumulant generating function (CGF) for heat exchange, thus
fully characterizing energy exchange in steady state.
The FCS treatment is appealing for several reasons: 
(i) It allows validation of the fluctuation symmetry, thus ensuring 
the thermodynamical consistency of approximate techniques. 
(ii) An FCS treatment automatically hands over the averaged current and high order cumulants. 
(iii) Practically, extending the Markovian population dynamics, Eq. (\ref{eq:qme}), 
to report on the FCS of heat exchange
is quite straightforward, as was exemplified in numerous studies \cite{EspositoReview,Bijay-recon,HavaNJP}. 

To derive the characteristic function for heat exchange, 
we follow the two-time measurement protocol \cite{EspositoReview} and introduce a counting parameter $\chi$ 
which tracks thermal energy entering/leaving the cold reservoir.
According to our sign convention, heat exchange (heat current) is positive when it flows towards the central system.  
In principle, one could keep track of energy transfer at each  boundary with the system
by introducing a counting parameter $\chi_\mu$ for each reservoir $\mu$ \cite{HavaNJP}. 
However, to simplify our analysis we limit it to a single counting parameter.

In the MQME approach, the FCS is obtained by writing down the 
counting field-dependent quantum master equation,
$\dot p(t,\chi) = {\cal L(\chi)} \, p(t,\chi)$,
with counting-field dependent Liouvillian and population.
The microscopic derivation of this equation was detailed e.g. in Ref. \cite{HavaNJP}, and exemplified in 
Appendix A and B for two- and three-level models, respectively.

The characteristic function for heat exchange is given by
%
${\cal Z}(t, \chi) =  \langle I| p(t,\chi)\rangle$,
where $\langle I | =(1, 1, 1,...)^T$ is the identity vector. 
The CGF for steady state heat exchange is obtained in the long time limit as
\bea
{\cal G} (\chi) = \lim_{t \to \infty} \frac{1}{t} \ln {\cal Z} (t,\chi).
\eea
Differentiating this function $k$ times with respect to $(i\chi)$ brings the 
$k$th cumulant of the process. 
Specifically, the first cumulant is the energy current between the system and the reservoir where counting is performed,
\bea
\langle J \rangle = \frac{\partial \cal G}{\partial(i\chi) } \Big|_{\chi=0}.
\label{eq:Jdef}
\eea
The second cumulant is the noise of that current. 

We obtain the CGF by diagonalizing ${\cal L}(\chi)$ 
and selecting the eigenvalue corresponding to the long-time limit. 
This is the eigenvalue with the smallest magnitude for its real part, 
$\lambda_{\textrm{min. real magnitude}}={\cal G}(\chi)$.
Formally, we solve the eigenvalue problem by writing down 
the characteristic polynomial for the matrix ${\cal L}(\chi)$
in terms of its eigenvalues, $\lambda_{1,2,...,N}$, 
\bea
\lambda_i^N + a_1(\chi) \, \lambda_i^{N-1} + ... + a_{N-1}(\chi) \, \lambda_i + a_N(\chi) = 0.
\nonumber\\
\label{eq:polynomial}
\eea 
The coefficients $a_j(\chi)$ are functions of the counting parameter.
In what follows, we distinguish between $a_j(\chi)$ and $a_j(0)$,
which are the coefficients of the
characteristic polynomials 
of ${\cal L}(\chi)$ and ${\cal L}(0) = {\cal L}$, respectively.
We now list several important properties of these elements.

The coefficients of the characteristic polynomial can be expressed in terms of the eigenvalues: 
 $a_j(\chi)$ of $\lambda^{N-j}$ is  given by the sum of all products of $j$
eigenvalues \cite{Brooks}, 
\bea
a_j(\chi)= (-1)^j \sum_{{\rm all\, sets\, of\,}  j\, \lambda's}  \lambda_k \times \lambda_p\times ... \times \lambda_q.
\label{eq:lambrule}
\eea  
Specifically,
\bea 
a_1(\chi) &=& -\trace [ {\cal L(\chi)}], 
\nonumber\\
a_N(\chi) &=& (-1)^N\det [ {\cal L}(\chi) ], 
\nonumber\\
 a_N(0)&=&0.
 \label{eq:math}
\eea
For the physical problems that we consider, the trace, $a_1(\chi)$ does not depend on the counting field, 
see  Appendices 
A and B. 
As for the other coefficients, from the general rule we get e.g. that for a 
$3\times 3$ matrix, $a_2(\chi)=\lambda_1\lambda_2 + \lambda_1\lambda_3+ \lambda_2\lambda_3$.
The last property in Eq. (\ref{eq:math}) stems from the fact that under our dynamics, the system
reaches a unique steady state, thus ${\cal L}(0)$ must acquire a zero eigenvalue, 
which nullifies the determinant.

From the detailed balance condition it can be shown that 
the eigenvalues of ${\cal L}(0)$  must be real 
since the eigenvalue problem can be mapped to a real symmetric matrix.
Furthermore, positive values for the eigenvalues are unphysical since that would lead
to exponentially growing probabilities.  \cite{ODEbook,OQSbook}.
Therefore, the eigenvalues of the rate matrix ${\cal L}(0)$ are real, negative numbers.
With that in mind, based on Eq. (\ref{eq:lambrule}), we conclude that the coefficients $a_{1,2,...N}(0)$ 
must be positive.
In particular, we later use the fact that $a_{N-1}(0)>0$.  

\subsection{Current and cooling condition}

For an $N\times N$ problem, ${\mathcal L}(\chi)$ has $N$ eigenvalues,
with ${\cal G}(\chi)$ as the  eigenvalue with the smallest-magnitude real part. 
As discussed in Ref. \cite{Dvira2018}, the eigenvalue problem can be greatly simplified if we focus on
the heat current only, Eq. (\ref{eq:Jdef}):
Because ${\cal G}(\chi=0) = 0$,  Eq. (\ref{eq:polynomial}) can be truncated 
to first order in $\lambda$ -- 
and would  still solve exactly for the first cumulant. 
More generally, Eq. (\ref{eq:polynomial}) can be truncated to the order of the desired cumulant. 

Truncating the characteristic polynomial (\ref{eq:polynomial}) to first order we get
\bea
{\cal G}^{(1)}(\chi) = -\frac{a_N(\chi)}{a_{N-1}(\chi)}. 
\label{eq:CGF}
\eea
To obtain the heat current we differentiate Eq. (\ref{eq:CGF}) with respect to $(i\chi)$.
Using $a_N(0)=0$, we arrive at
\bea
\langle J \rangle =  \frac{\partial {\cal G}^{(1)}(\chi)}{\partial (i\chi)} \Big|_{\chi=0} 
= -\frac{1}{a_{N-1}(0)} \frac{\partial a_N(\chi)}{\partial (i\chi)} \Big|_{\chi=0}.
\label{eq:CGFJ}
\eea
%

We recall that $a_N(\chi)= (-1)^N \textrm{det} \left[ {\cal L}(\chi) \right]$. 
Applying Jacobi's formula, which connects the derivative of the determinant to a trace relation, 
we get an analytic expression for the heat current, from the reservoir where counting is performed
towards the quantum system,
\bea
\langle J \rangle = \frac{(-1)^{N+1}}{a_{N-1}(0)}  \trace \left[ \textrm{adj}\left[ {\cal L}(0) \right] \ \frac{\partial {\cal L}(\chi)}{\partial (i\chi)} \Bigg|_{\chi=0}  \right].
\label{eq:current} 
\eea
Here, $\textrm{adj}\left[ {\cal L}\right]$ is the adjugate matrix of the Liouvillian; the transpose of the cofactor matrix of ${\cal L}$.
%
If counting is performed on the cold bath, Eq. (\ref{eq:current}) provides
the cooling current $\langle J_C \rangle$. However, the equation can be readily applied to calculate other currents 
by assigning the counting parameter to the respective heat source.

We now write down a general condition on the cooling window.
Since $a_{N-1}(0)>0$,  cooling is achieved if
\bea
(-1)^{N+1}  \trace \left[ \textrm{adj}\left[ {\cal L}(0)\right] \ \frac{\partial {\cal L}(\chi)}{\partial (i\chi)} 
\Bigg|_{\chi=0}  \right] > 0.
\label{eq:cooling}
\eea
Equations (\ref{eq:current})-(\ref{eq:cooling}) are the main results of this paper.
They offer the following benefits over standard calculations of the heat current \cite{commentJ}: 

(i) Equation  (\ref{eq:current}) allows us to feasibly calculate heat currents 
{\it analytically} for systems described by the generic evolution, Eq. (\ref{eq:qme}), without a prior knowledge of
the steady state population. 
Similarly, the cooling condition  (\ref{eq:cooling})  can be evaluated analytically
to easily determine if a quantum system is suitable to act as a refrigerator.

(ii) In both Equations (\ref{eq:current}) and (\ref{eq:cooling}) only the diagonal elements of the 
matrix product are required to perform the trace operation.
The matrix $\partial {\cal L}(\chi)/\partial (i\chi)|_{\chi=0}$ is sparse in typical models,
thus drastically reducing the complexity of the calculation.
Importantly, this matrix can be readily-intuitively constructed, and it does not require knowledge of the principles
of the FCS formalism:
As we illustrate  in the Appendices, when calculating $\langle J_{\nu}\rangle$
the terms surviving in $\partial {\cal L}(\chi)/\partial (i\chi)|_{\chi=0}$
are energy transfer rates to/from the reservoir $\nu$ and the system.

(iii) 
Eq. (\ref{eq:current}) can be used to study heat exchange at each contact,
thus providing the cooling coefficient of performance, that is the
ratio of cooling current to input power extracted from the work reservoir.

(iv) Our derivation for the cooling current does not discriminate between different sources of 
irreversibility: whether these are due to the competition between different cooling pathways 
(``internal dissipation") or due to heat leaks with multiple reservoirs
coupled to the driving (work) transition. 

(v) Our derivation
does not  rely on the additivity of the total Liouvillian
with the different baths. Therefore, expressions (\ref{eq:current})-(\ref{eq:cooling}) can be applied to more general setups,
such as the non-additive system-bath models discussed in Ref. \cite{HavaNJP}.

Appendix A exemplifies the evaluation of the heat current in a two-level two-bath setup.
In Appendix B, we construct three-level three-bath models and illustrate the calculation of the
cooling current and cooling condition in ideal and non-ideal QARs.

%

\subsection{Noise power}

It is possible to obtain the noise power of a specific current, 
$\langle S \rangle =  \frac{\partial^2 \cal G}{\partial(i\chi)^2}\Big|_{\chi=0}$ 
by following a similar procedure to the calculation of the heat current.
To achieve that,
we use a characteristic polynomial truncated to second order in $\lambda$,
$a_{N-2}(\chi)\lambda^2 + a_{N-1}(\chi)\lambda + a_N(\chi)=0$,
which yields the CGF,
\bea
{\cal G}^{(2)} (\chi)=  \frac{-a_{N-1}(\chi) + \sqrt{ a_{N-1}(\chi)^2 -4a_N(\chi)a_{N-2}(\chi) } }{2a_{N-2}(\chi)}.
\nonumber\\
\eea
The resulting noise expression [after a second derivative in ($i\chi$)]
is  cumbersome. Therefore, we only consider 
classes of models for which both $a_{N-1}(\chi)$ and $a_{N-2}(\chi)$ 
do not depend on the counting parameter. 
This is the case, for example, for a three level system
if we further assume that the cold bath is coupled 
to a specific transition that no other bath is coupled to \cite{commentN3}.  
Under this approximation, the noise can be received via 
\bea
\langle S \rangle = \frac{(-1)^{N+1}}{ a_{N-1}(0)}   &\trace& \left[ 
\frac{\partial \textrm{adj} [{\cal L(\chi)}] }{\partial (i\chi) }
\frac{\partial {\cal L} }{\partial (i\chi) }  
+ \textrm{adj} [{\cal L}(0)]
 \frac{\partial^2 {\cal L} }{\partial (i\chi) ^2} \right] _{\chi=0} \nonumber \\
&-&2 \left(\frac{a_{N-2}(0)}{ a_{N-1}(0)} \right) \langle J \rangle^2.
\label{eq:noise} 
\eea
For a given device setup, the noise could be minimized with respect to tunable parameters 
to determine the fundamental limit of noise reduction. 
While this expression seems cumbersome,
the extreme sparsity of the differentiated Liouvillian matrices  
drastically simplifies calculations.

\section{Case study: Three-level QAR with competing cycles and leaks} 
\label{sec:3LQAR}

A QAR is a continuous-cycle heat machine, which operates without external driving.
It extracts energy from a cold ($C$) reservoir assisted by heat supplied from a so-called work reservoir ($W$),
and it dumps the heat into a hot reservoir ($H$).
The simplest version of such a device, the three-level quantum absorption refrigerator \cite{Scovil-Schluz-Dubois},
has been explored in detail in the weak system-bath coupling limit \cite{Kosloff2001,Skrzypczyk2010}.

We exemplify the analytic results derived in Sec. (\ref{sec:cooling}) on a general, non-ideal three-level QAR.
The setup includes three levels, $\ket{1}$, $\ket{2}$, and $\ket{3}$.
Transitions between states $\ket{i}$ and $\ket{j}$ are induced by three heat baths,
with $\beta_C >\beta_H>\beta_W$; $\beta_{\mu}$ is the inverse temperature of the $\mu$ bath.
The Hamiltonian of the setup is given by,
\bea
\hat H = \sum_{j=1}^3 E_j \ket{j}\bra{j} 
+ \sum_{\substack {\mu = C,H,W}} \sum_{k}\omega_{\mu,k} \, \hat a^{\dagger}_{\mu,k} \hat a_{\mu,k} 
\nonumber \\
+ \sum_{\substack {\mu = C,H,W}} \sum_{\substack {j > i}}  \hat B_{ij}^{\mu}(\ket{i}\bra{j} + h.c.)
\nonumber\\
\label{eq:H}
\eea
Here, $\hat a^{\dagger}_{\mu,k}$ ($\hat a_{\mu,k}$) is a bosonic creation (annihilation) operator
of mode $k$ in the $\mu$ bath. 
$\hat B_{ij}^{\mu}=\sum_ku_{ij}^{\mu,k} \left( \hat a^{\dagger}_{\mu,k} + \hat a_{\mu,k}\right)$
is a bath operator that couples to the $i\leftrightarrow j$ transition
with energy $u_{ij}^\mu$, which is assumed to be a real number.

The Hamiltonian (\ref{eq:H}) is a general version of the ideal 3-level quantum absorption refrigerator, in which
only the  $u_{12}^C$, $u_{23}^W$ and $u_{13}^H$ transition elements are nonzero;
 this scenario is `ideal' because  tuning the 
 system's parameters (energy levels) 
 can achieve the Carnot cooling limit, see Appendix B and e.g. Refs. \cite{Correa2014Nature,Dvira2018}. 

\begin{figure*}[t]
\begin{subfigure}{.5\textwidth}
  \centering
  \includegraphics[width=.8\linewidth]{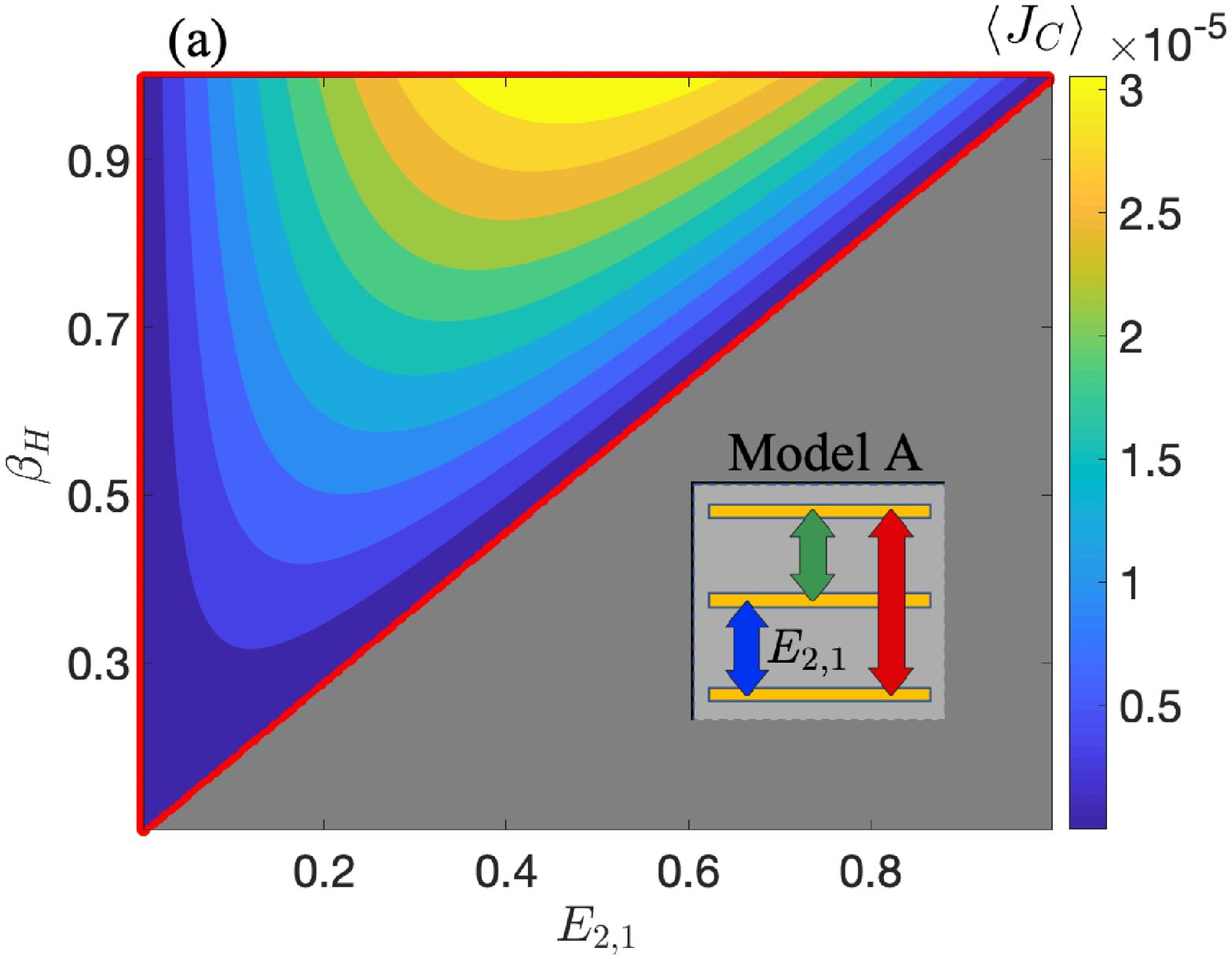}
  \label{fig:sfig1}
\end{subfigure}%
\begin{subfigure}{.5\textwidth}
  \centering
  \includegraphics[width=.8\linewidth]{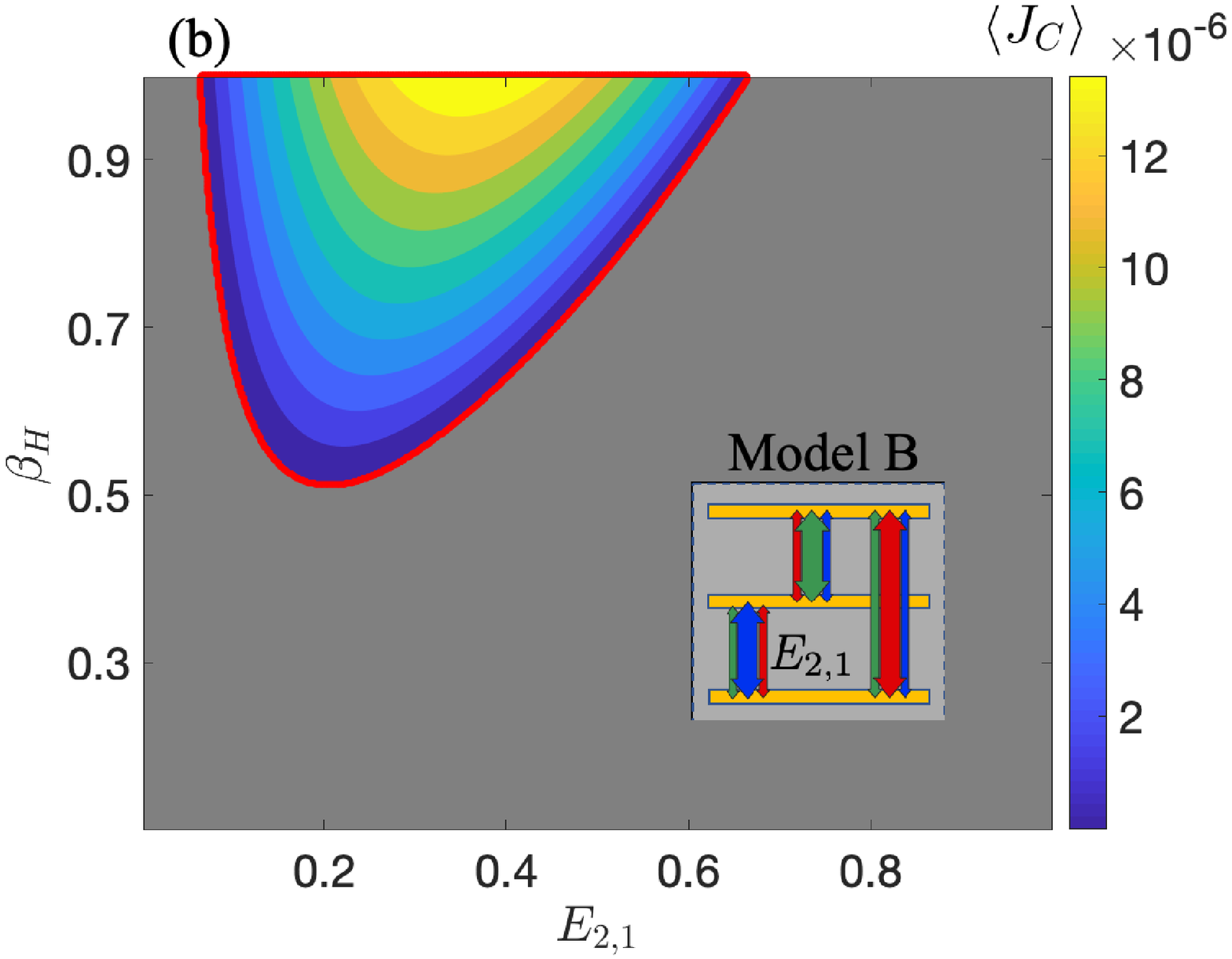}
  \label{fig:sfig2}
\end{subfigure}
\begin{subfigure}{.5\textwidth}
  \centering
  \includegraphics[width=.8\linewidth]{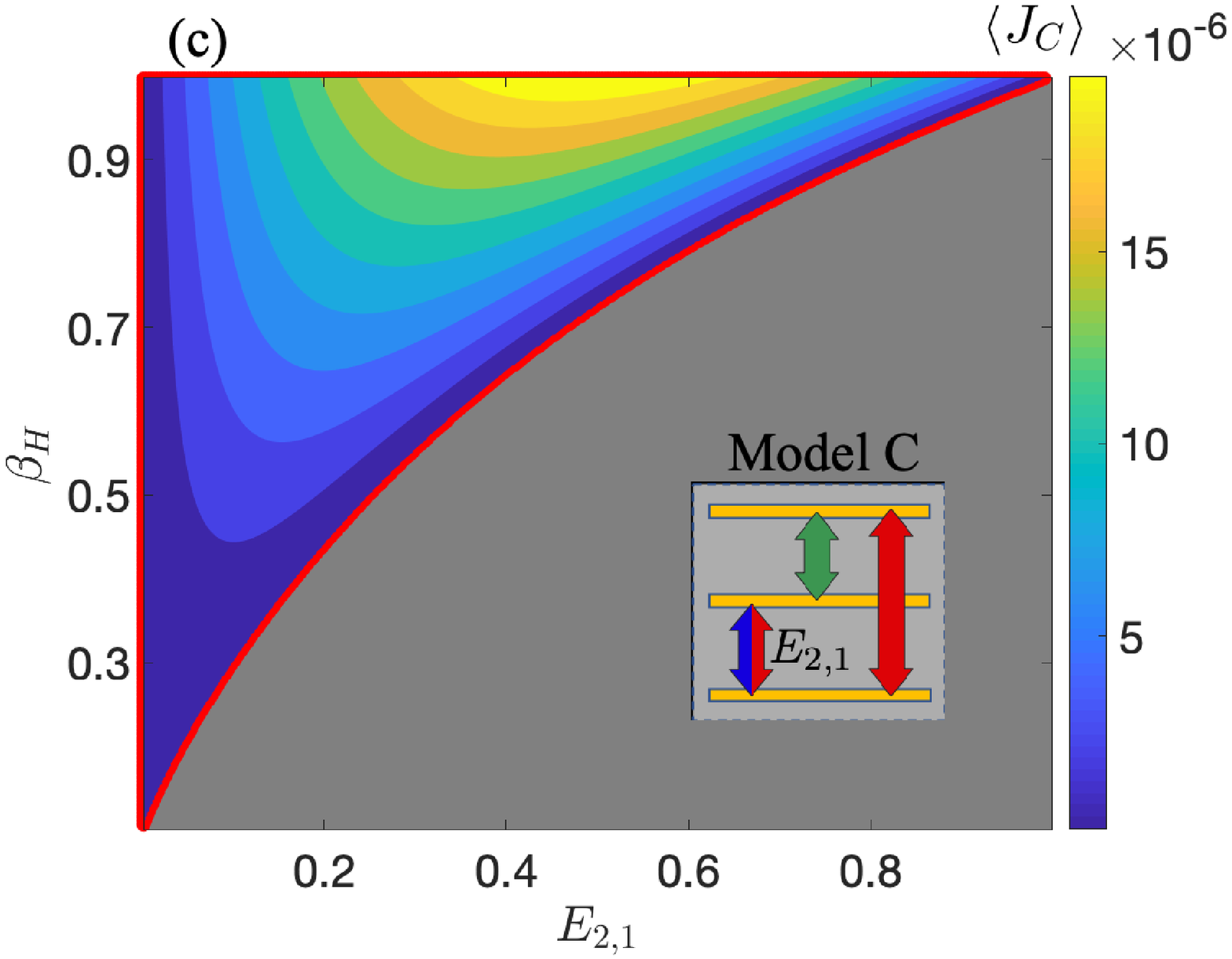}
  \label{fig:sfig3}
\end{subfigure}%
\begin{subfigure}{.5\textwidth}
  \centering
  \includegraphics[width=.8\linewidth]{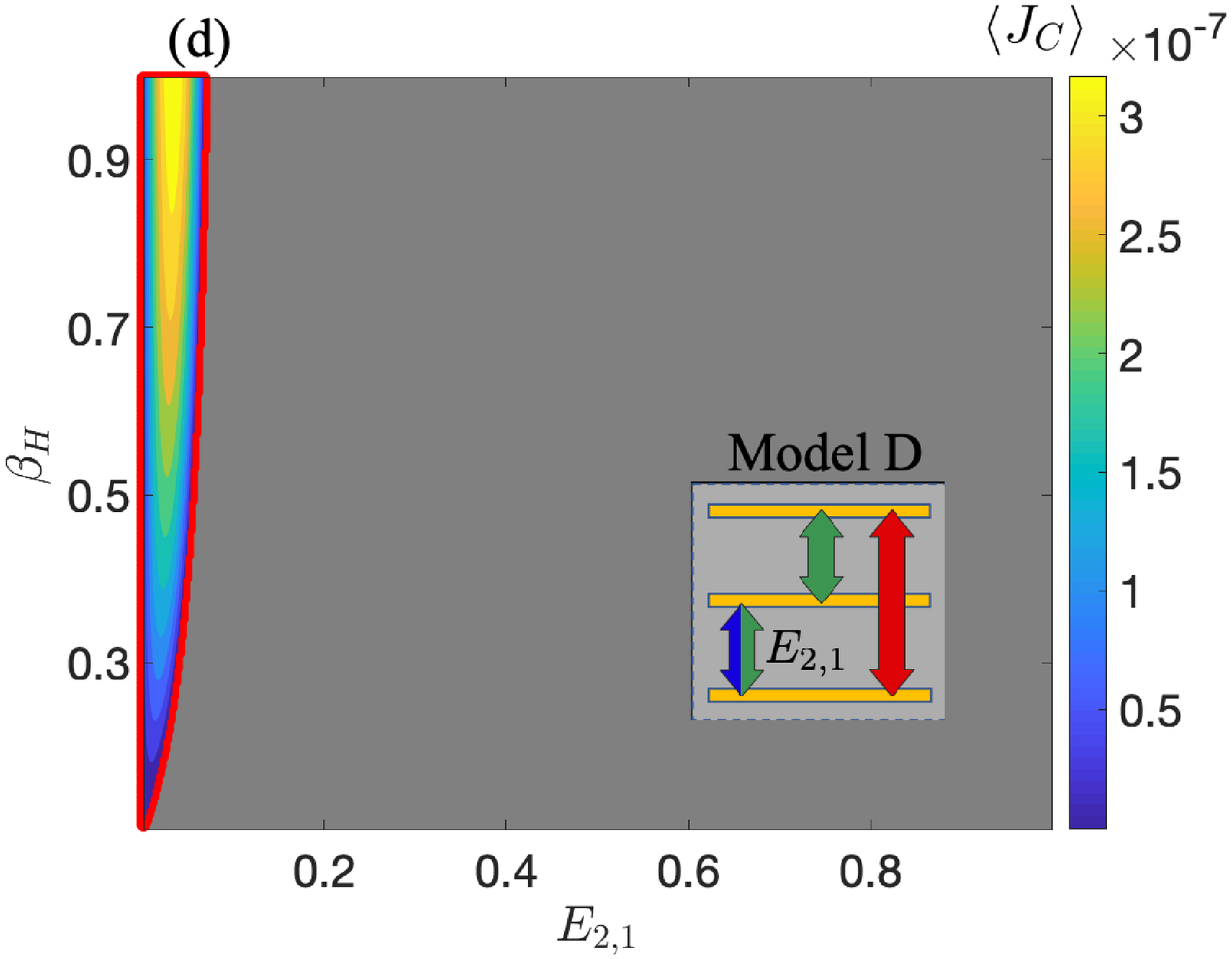}
  \label{fig:sfig4}
\end{subfigure}
\caption{
Cooling currents of three-level QAR as a function of energy spacing $E_{2,1}$ 
and the inverse temperature of the hot reservoir, $\beta_H$.
The colored region corresponds to a positive current: heat extracted from the cold bath.
The red outline encloses the region where Eq. (\ref{eq:cooling}) predicts cooling to occur. 
Parameters are $E_{3,1}=1$, $\beta_C=1$, $\beta_W=0.1$, $\omega_C=10$, $\gamma=0.001$ and 
$\tilde \gamma=\gamma/50$.
(a) Model A. Ideal QAR setup, $\gamma=\gamma^C_{12}=\gamma^H_{13}=\gamma^W_{23}$.
(b) Model B. QAR with dominant transitions, $\gamma=\gamma^C_{12}=\gamma^H_{13}=\gamma^W_{23}$ and 
additional weaker transitions,  $\tilde \gamma=\gamma^{C,W}_{13}=\gamma^{C,H}_{23}=\gamma^{W,H}_{12}$.
(c) Model C. QAR with heat leaks from the hot bath, 
$\gamma=\gamma^{C,H}_{12}=\gamma^H_{13}=\gamma^W_{13}$.  
(d)  Model D. QAR with heat leaks from the work bath, 
$\gamma=\gamma^{C,W}_{12}=\gamma^H_{13}=\gamma^W_{23}$. 
Coupling coefficients that are not indicated are null.
}
\label{fig:contours}
\end{figure*}


In what follows, we study the cooling performance of non-ideal setups.
Specifically, in our work, 
the term `heat leaks' refers to having several baths coupled to the same transition. 
This leads to a competition between the different baths, to either excite or relax population. 
Internal dissipation arrises when cooling can be achieved by more than one cycle \cite{Correa2015PRE}.

We can feasibly predict the parameters for the cooling window by writing down the Liouvillian
${\cal L}(\chi) = {\cal L}_C(\chi)+{\cal L}_W+{\cal L}_H$. 
For harmonic baths and bilinear system-bath couplings the 
Liouvillians are made from the rate constants,
\bea
k_{i\to j}^{\mu}= \begin{cases}
\Gamma^{\mu}_{ij}(E_{j,i})n_{\mu}(E_{j,i}) & E_j>E_i\\
\Gamma^{\mu}_{ij}(E_{i,j})[n_{\mu}(E_{i,j})+1] & E_j<E_i. 
\label{eq:kij}
\end{cases}
\eea
%
Here, $n_{\mu}(\omega)=[e^{\beta_{\mu}\omega}-1]^{-1}$ is the Bose-Einstein occupation function.
$\Gamma^{\mu}_{ij}(\omega)=2\pi\sum_{k} (u_{ij}^{\mu,k})^2\delta(\omega-\omega_{\mu,k} )$ is the 
spectral density function of the baths, $E_{i,j}=E_i-E_j$.
Specifically, we employ an ohmic model,
 $\Gamma^{\mu}_{ij}(\omega) = \gamma_{ij}^{\mu}\omega e^{-|\omega|/\omega_c}$ with
$\gamma$  as a dimensionless coupling parameter and $\omega_c $ the cutoff frequency, assumed to be high.
Since we study the problem in the weak-coupling limit, the full functional form of the spectral function carries 
no impact on the cooling current;
only the value in resonance with the system's transitions is used in calculations.

We examine different configurations of the QAR model in the four panels of Fig. (\ref{fig:contours}).
In each case, we indicate schematically the allowed transitions as activated by the different baths using colored arrows:
blue for the $C$ bath, red for $H$, and green for $W$.
The relative width of the arrows indicate the relative strength of that bath-induced transition. 
The red curve encloses the region where cooling is predicted according to Eq. (\ref{eq:cooling}), 
and the contours show the positive cooling current calculated numerically.  
For visibility, negative current (heat flow directed towards the cold bath) is not presented (gray region).
As can be seen, the cooling condition (red curve) perfectly predicts the cooling window for each of the QAR setups.

Our analysis demonstrates that the current at the cold contact can be generally written as  
\bea
\langle J_C\rangle = \sum_{i>j}F_{cyc}^{i,j} + \sum_{\mu, i>j}F_{leak,\mu}^{i,j}.
\label{eq:QARG}
\eea
Here, $F_{cyc}^{i,j}$ identifies a circuit that can realize a QAR
with the cold bath coupled to the $i\leftrightarrow j$ transition and the work and hot  baths completing a cycle, e.g.
$i\xrightarrow{C} j \xrightarrow{W}j' \xrightarrow{H}i$, and the reversed process. 
`Cycle' contributions must involve the three baths.
The second term, $F_{leak,\mu}^{i,j}$,  is always negative. 
It comprises heat leaks from the $\mu$ bath to the cold bath
when both reservoirs are coupled to the $i\leftrightarrow j$ transition.  
Thus, leak terms involve two reservoirs, either $H$ or $W$, and $C$. 
This classification (see Appendix B) agrees with \cite{AlonsoNJP17}.
We now use Eq. (\ref{eq:QARG}) to explain the results of Fig. \ref{fig:contours}.

Model A. The ideal QAR setup is constructed by taking nonzero values for 
$\gamma_{12}^C$, $\gamma_{23}^W$, $\gamma_{13}^H$;
all other couplings are null.
As we show in panel a, this setup achieves the widest cooling window and the most substantial cooling current. 
We note that there is an optimal level spacing $E_{2,1}$ for maximizing the cooling current, 
which depends on the temperature of the cold bath. This setup is termed `ideal' since it can achieve the 
Carnot bound.
Increasing $\beta_H$ (decreasing $T_H$) is beneficial for cooling:
When the temperature gradient between $H$ and $C$ is reduced, heat flow from $H$ to $C$ is minimized.
In Appendix B we use Eq. (\ref{eq:cooling}) and derive the cooling condition, $F_{cyc}^{2,1} >0$, or explicitly
\bea
F_{cyc}^{2,1}&=& E_{2,1}k_{3\to 1}^H k_{3\to 2}^Wk_{2\to 1}^C 
\nonumber\\
&\times &
\left(e^{-\beta_W E_{3,2}} e^{-\beta_C E_{2,1}} - e^{-\beta_H E_{3,1}}\right)>0.
\label{eq:CI}
\eea
It agrees with previous studies, see e.g. \cite{Correa2014Nature}.
Next, the ideal configuration is compromised by including additional couplings, resulting in heat leaks.

Model B.  Here we distinguish between 
`dominant' couplings, which reproduce the ideal setup,
$\gamma=\gamma_{12}^{C}=\gamma_{23}^{W}=\gamma_{13}^{H}$,
and secondary-weaker couplings, which lead to heat leakage and competition between cycles,
$\tilde \gamma=\gamma_{13}^{C,W}=\gamma_{23}^{C,H}=\gamma_{12}^{H,W}$, with
$\tilde \gamma\ll \gamma$.
Results are presented in panel b. 
We find that the cooling function disappears for small $E_{2,1}$, but it survives in the intermediate $E_{2,1}\sim E_{3,2}$ regime.
We can rationalize our observations as follows (Appendix B):
While heat directly leaks to the cold bath from both the hot and work reservoirs,
 we may extract energy from the cold bath using the different cycles.
We note that $F_{cyc}^{3,1}$ is always negative 
thus we are left with two cycles, $F_{cyc}^{2,1}$ and $F_{cyc}^{3,2}$.
The cooling condition for cycle $F_{cyc}^{2,1}$ is
\bea
\frac{E_{2,1}}{E_{3,1}} \leq \frac{\beta_H-\beta_W}{\beta_C-\beta_W}.
\label{eq:condc1}
\eea
In contrast, cooling via the cycle $F_{cyc}^{3,2}$ is achieved if we satisfy
\bea
\frac{E_{2,1}}{E_{3,1}} \geq \frac{\beta_C-\beta_H}{\beta_C-\beta_W}.
\label{eq:condc2}
\eea
The two cycles have conflicting requirements on the energy gap $E_{2,1}$, thus cooling is not achievable for
either small or large $E_{2,1}$, when one of the cycles delivers heat into the cold bath.
Furthermore, heat leaks and the non-cooling $F_{cyc}^{3,1}$ cycle suppress the cooling current for both small and large $E_{2,1}$
spacing.

Model C. 
In this model (panel c) the hot  bath directly competes with the cold bath on the cooling performance. 
As we show in Appendix B, the cooling condition of the setup can be compactly written as
\bea
F_{cyc}^{2,1} + F_{leak,H}^{2,1} >0,
\label{eq:modelC}
\eea
with $F_{leak,H}^{2,1}\propto\left( e^{-\beta_{C}E_{2,1}} - e^{-\beta_{H}E_{2,1}}\right)$, which is negative.
Therefore,  the extracted current is smaller than in Model A, 
and the cooling condition is more difficult to satisfy, 
resulting in a more limited
window of operation.
Specifically, beyond a certain value for $E_{2,1}$, 
the cooling function is lost 
since the 
heat leak term takes over. 

Model D. This setup (panel d) is similar to Model C, but
with the work bath directly competing with the cold bath. 
A cooling condition, analogous to Eq. (\ref{eq:modelC}), can be
written. However, given that $\beta_W\ll \beta_C$, 
the cooling performance is significantly reduced, and it survives only for very small energy splitting, $E_{2,1}$.

The secular approximation breaks down when two quasidegenerate levels
are coupled to a third level, with the same bath responsible for transitions between the ground and excited states.
In Fig. \ref{fig:contours}, the energy level $E_{2}$ is allowed to move 
in the full range between $E_1$ and $E_3$, thus reaching degeneracy 
in the extreme limits. However, our analysis and conclusions do not pertain to these special limits.
Specifically, in panel b the region of no-cooling at small $E_{2,1}$ is controlled by $\tilde \gamma$, 
and not by coherent effects.
In panel d, our main observation is the suppression of the central cooling region 
(when $E_2$ is set about halfway in the $E_{3,1}$ gap). 
This suppression is controlled by a leakage process from the work bath.

\begin{figure}
  \includegraphics[width=.8\linewidth]{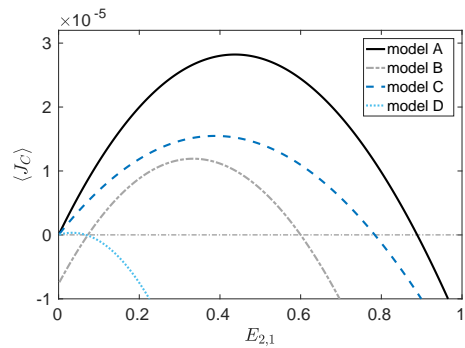}
\caption{
Heat current $\langle J_C \rangle$  with varying level spacing, $E_{2,1}$. 
The different curves correspond to the QAR results as presented in Fig. \ref{fig:contours} for
$\beta_H=0.9$.
}
\label{fig:current-noise}
\end{figure}

In Fig. \ref{fig:current-noise} we display examples based on Fig. \ref{fig:contours}
and highlight the different magnitudes of the cooling currents.
Because $T_H$ is close to $T_C$, the cooling behavior is moderately-negatively impacted in models B and C relative
to model A. In contrast, heat leaks from the work to the cold reservoir (model D)
dramatically lessen both the cooling window and the cooling current.

\section{Summary}
\label{sec:summ}

We derived a cooling condition for a multilevel quantum absorption refrigerator described at the level
of a Markovian quantum master equation. The obtained cooling condition is given in terms of the generator of the dynamics, 
and it provides analytical insight to the machine operation. 
It allows us to determine the effect of heat leaks and competing cycles in the setup, 
and feasibly identify the parameter space that achieves the desired function. 
We further obtained a closed-form expression for the currents (cooling or input power) in the model,
and described the calculation of the current noise.
We expect this work to assist in the quest to realize QARs; 
a realistic device evidently suffers from leakage and internal dissipation processes.
While we exemplified our results on a three-level QAR, the formalism holds for a general $N$-level system,
 with the heat current combining cycles and leaks as in Eq.  (\ref{eq:QARG}).

Steady-state coherences arise in $V$ and $\Lambda$-type level structure when nonequilibrium baths are coupled 
to the energy degenerate transitions \cite{Michael2018}. 
The impact of such coherences on the cooling current 
can be studied numerically with a non-secular MQME \cite{Michael2018,Noise2018,Pet,Zambrini}. 
A recent study showed that under some assumptions, 
quantum coherent devices can be mapped onto a ``classical emulator" that reproduces the same 
thermodynamic performance in the long time limit \cite{Correa19PRE}, 
where our formalism could be applied. 
Achieving a compact, feasible cooling condition in cases that cannot be described by a classical-incoherent emulator
is left for future work.

\begin{acknowledgments}
DS acknowledges support from an NSERC Discovery Grant and the Canada Research Chair program.
The work of HMF was supported by the NSERC PGS-D program, 
David H. Farrar Graduate Scholarship in Chemistry, and the Lachlan Gilchrist Fellowship Fund. 
\end{acknowledgments}

\renewcommand{\theequation}{A\arabic{equation}}
\setcounter{equation}{0}  
\section*{Appendix A:
The nonequilibrium spin-boson model}

\begin{figure}[H]
\centering
\includegraphics[width=.8\linewidth]{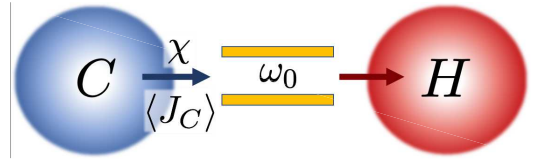}
\caption{
Scheme of the nonequilibrium two-level system. By introducing a counting parameter on the cold bath, 
we calculate the heat current extracted from this bath to the system, which is negative in this two-terminal problem. 
}
\label{fig:TLS}
\end{figure}

We show that Eqs. (\ref{eq:current}) and (\ref{eq:noise}) reproduce known results 
for the heat current and its noise in the nonequilibrium spin boson model ($N=2$), 
displayed in Fig. (\ref{fig:TLS}).  The Hamiltonian of the model is 
\bea
\hat{H} = \frac{\omega_0}{2} \hat  \sigma_z + \sum_{\substack {\mu,k}} \omega_{\mu,k} \, \hat a^{\dagger}_{\mu,k} \hat a_{\mu,k},  
+ \hat \sigma_x \sum_{\mu} \hat B^\mu,
\eea
where $\hat B^\mu = \sum u^{\mu,k} (\hat a_{\mu,k}^\dagger+\hat a_{\mu,k})$. 
The counting field dependent MQME reads 
${\dot p(t,\chi) }= {\cal L}(\chi) {p(t,\chi)}$ with
the Liouvillian rate matrix  \cite{HavaNJP}, 
\bea
{\cal L}(\chi) = 
\begin{pmatrix} 
-k_{1 \to 2}^C - k_{1 \to 2}^H 					& k_{2 \to 1}^C e^{-i\chi \omega_0} + k_{2 \to 1}^H \\
k_{1 \to 2}^C e^{i\chi \omega_0} + k_{1 \to 2}^H 	& -k_{2 \to 1}^C - k_{2 \to 1}^H 
\end{pmatrix}
\nonumber
\eea
In this equation, rate constants describing heat exchange with the cold bath are decorated by the counting fields.
The `bare' rate constants are 
$k_{1 \to 2}^\mu = \int_{-\infty}^{\infty} dt \, e^{-i \omega_0 t} \langle \hat B^\mu (t)  \hat B^\mu (0)  \rangle $, 
and the detailed balance condition enforces $k_{1\to 2}^\mu = e^{-\beta_\mu \omega_0} k_{2 \to 1}^\mu $. 
To use equation (\ref{eq:current}) we calculate the derivative,
\bea
\frac{\partial {\cal L}(\chi)}{\partial (i \chi)} \Bigg|_{\chi=0} = 
\begin{pmatrix} 
0				& -\omega_0 \, k_{2 \to 1}^C  \\
\omega_0 \, k_{1 \to 2}^C  	& 0
\end{pmatrix}. \nonumber
\eea
We note the natural form of this matrix, allowing us to create it intuitively:
It includes heat transfer rates to/from the cold bath with the appropriate sign convention. 

Using Eq. (\ref{eq:current})  we  can now determine the current from the $C$ bath, 
\bea
\langle J_C \rangle = \frac{-1}{a_1(0)} \trace \left[ 
\begin{pmatrix} 
C_{11} & C_{21}  \\
C_{12} & C_{22}
\end{pmatrix}
\begin{pmatrix} 
0				& -\omega_0 \, k_{2 \to 1}^C  \\
\omega_0 \, k_{1 \to 2}^C  	& 0
\end{pmatrix} \right] \nonumber
\eea
The matrix filled with the cofactor terms $C_{ij}$ is the adjugate matrix, $\textrm{adj}[{\cal L}(0)]$. 
$a_1(0)$ is the coefficient of $\lambda$ in the characteristic polynomial, Eq. (\ref{eq:polynomial}).
Simplifying, and marking only the relevant terms, while using the fact that $a_1(0)=\trace\left[{\cal L}(0)\right]$, 
we get
\bea
\langle J_C \rangle = \frac{-1}{\trace\left[{\cal L}(0)\right]} \trace \left[ 
\begin{pmatrix} 
C_{21} \omega_0 \, k_{1 \to 2}^C & ...  \\
...  	& -C_{12} \omega_0 \, k_{2 \to 1}^C
\end{pmatrix} \right] . \nonumber\\
\label{eq:JTLS1}
\eea
We need only solve for $C_{12}$ and $C_{21}$ from the adjugate matrix. 
Trivially, $C_{12}=- [\mathcal L(0)]_{2,1}= 
(k_{1\to 2}^C + k_{1\to 2}^H)$ and
$C_{21}=- [\mathcal L(0)]_{1,2}= 
(k_{2\to 1}^C + k_{2\to1}^H)$, which results in 
\bea
\langle J_C \rangle = \frac{\omega_0 \left[ 
k_{1 \to 2}^H  k_{2 \to 1}^C - k_{2 \to 1}^H  k_{1 \to 2}^C  
 \right]}
{k_{1 \to 2}^C + k_{1 \to 2}^H+k_{2 \to 1}^C +k_{2 \to 1}^H}.
\eea
Using the harmonic bath model and bilinear system-bath coupling the rate constants are
$k_{2 \to 1}^\mu = \Gamma_{\mu}(\omega_0) [n_\mu(\omega_0)+1]$ 
and $k_{1 \to 2}^\mu = \Gamma_\mu(\omega_0) n_\mu(\omega_0)$, therefore
\bea
\langle J_C \rangle = \frac{\omega_0 \Gamma_C(\omega_0) \Gamma_H (\omega_0) \left[ 
n_C(\omega_0)-n_H(\omega_0)\right]}
{\Gamma_C(\omega_0)[1+2n_C(\omega_0)] + \Gamma_H(\omega_0)[1+2n_H(\omega_0)]}.
\label{eq:JTLS2}
\nonumber\\
\eea
Here, $\Gamma_{\mu}(\omega_0)=2\pi\sum_{k}(u^{\mu,k})^2\delta(\omega_0-\omega_{\mu,k})$  
is the spectral density of the $\mu$ bath evaluated at $\omega_0$ 
and $n_\mu(\omega_0)=(e^{\beta_\mu \omega_0}-1)^{-1}$ 
is the Bose-Einstein distribution function. 

Equation (\ref{eq:JTLS2}) agrees with previous studies \cite{Segal04,Segal06,Yelena10}.
However, the present evaluation, which is based on Eq. (\ref{eq:JTLS1}) is almost trivial compared to the standard approach,
which requires one to solve the equation of motion for the population in steady state, then substitute 
the population in the expression for the current.

Beyond the current, we follow the procedure for calculating the current noise 
using Eq. (\ref{eq:noise}). In this case we need to examine two matrices
$\partial^2 {\cal L(\chi)}/\partial (i\chi)^2$ 
and $\partial \textrm{adj}\left[{\cal L}(\chi)\right]/\partial (i\chi)$. 
The former is easy to calculate, 
\bea
\frac{\partial^2 {\cal L}(\chi)}{\partial (i \chi)^2} \Bigg|_{\chi=0} = 
\begin{pmatrix} 
0				& \omega_0^2 \, k_{2 \to 1}^C  \\
\omega_0^2 \, k_{1 \to 2}^C  	& 0
\end{pmatrix}. \nonumber
\eea
As for the latter, again, we only need to study two elements, 
\bea
\frac{\partial C_{21}}{\partial (i\chi)} = \omega_0 k^C_{2 \to 1}, \,\,\,\,\,\,\,\,\,\,
\frac{\partial C_{12}}{\partial (i\chi)} = -\omega_0 k^C_{1 \to 2}.\nonumber \\ 
\eea
Altogether, we get
\bea
\langle S_C \rangle &=& \frac{\omega_0^2 \left[ k_{2 \to 1}^H k_{1 \to 2}^C + k_{1 \to 2}^H k_{2 \to 1}^C \right] - 2 \langle J_C \rangle^2}{k_{1 \to 2}^C + k_{1 \to 2}^H+k_{2 \to 1}^C +k_{2 \to 1}^H},
\nonumber\\
\eea
which reduces to the known result \cite{Yelena10} 
\bea
\langle S_C \rangle = \frac{\omega_0^2 \, \Gamma_C\Gamma_H  \left[ 
(1+n_C)n_H+(1+n_H)n_C\right] - 2 \langle J_C \rangle^2}
{\Gamma_C(1+2n_C) + \Gamma_H(1+2n_H)}.
\nonumber\\
\eea
Here, $\Gamma_\mu$ and $n_\mu$  are evaluated at the frequency of the spin, $\omega_0$. 

\renewcommand{\theequation}{B\arabic{equation}}
\setcounter{equation}{0} 
\section*{Appendix B:
Ideal and non-ideal three-level QAR}
\label{appendix:ideal3}
We describe here the calculation of the cooling window in the different 3-level models presented in 
the four panels of Fig. \ref{fig:contours}.

\subsection{Ideal QAR: Model A}
We exemplify our procedure to obtain a cooling condition [Eq. (\ref{eq:cooling})] 
on the canonical three-level quantum absorption refrigerator
with `ideal' coupling operator. 
The Hamiltonian of the model reads
\bea
\hat{H} &=& \sum_{j=1}^3 E_j \ket{j}\bra{j} + \sum_{\substack {\mu,k}} \omega_{\mu,k} \, \hat a^{\dagger}_{\mu,k} \hat a_{\mu,k}   \nonumber \\%
&+& \hat B^C_{12} \left(\ket{1}\bra{2} + \ket{2}\bra{1}\right)
+ \hat B^H_{13} \left(\ket{1}\bra{3} + \ket{3}\bra{1}\right)  
\nonumber \\ &+& \hat B^W_{23} \left(\ket{2}\bra{3} + \ket{3}\bra{2}\right).
\eea
where $\hat B^\mu_{ij} = \sum_k u^{\mu,k}_{ij} (\hat a^\dagger_{\mu,k}+a_{\mu,k})$
is an operator of the $\mu$ heat bath.
The equations of motion for the populations are given by $\dot p(t) = {\cal L}(0) \, p(t)$
with the population vector $p(t)=(p_1(t), p_2(t), p_3(t))^T$ and the Liouvillian
%
%
%
\bea
{\cal L}(0) =
\begin{pmatrix}
-k_{1 \to 2}^C - k_{1 \to 3}^H           & k_{2 \to 1}^C   & k_{3 \to 1}^H\\
k_{1 \to 2}^C & -k_{2 \to 1}^C - k_{2 \to 3}^W  & k_{3 \to 2}^W\\
k_{1 \to 3}^H           &               k_{2 \to 3}^W           &       -k_{3 \to 1}^H - k_{3 \to 2}^W
\end{pmatrix}
\nonumber\\
\eea
The rate constants are
\bea
k^C_{1\to 2} &=& \Gamma_C(E_{2,1}) n_C(E_{2,1}), 
\nonumber \\
k_{1\to 3}^H &=& \Gamma_H(E_{3,1}) n_H(E_{3,1}),
\nonumber\\
k_{2\to 3}^W &=& \Gamma_W(E_{3,2}) n_W(E_{3,2}).
\eea
The detailed balance relation dictates the rate constants of reversed processes, e.g.,
$k_{2\to 1}^C =  \Gamma_C(E_{2,1}) \left[n_C(E_{2,1})+1\right]$.
The system-bath coupling constants (hybridization) are
$\Gamma_{\mu}(\omega)=2\pi \sum_{k}(u_{ij}^{\mu,k})^2\delta(\omega_{\mu,k}-\omega)$.  
The Bose-Einstein occupation factors are $n_{\mu}(\omega)=[e^{\beta_{\mu}\omega}-1]^{-1}$,
given in terms of the inverse temperature $\beta_{\mu}=1/(k_BT_{\mu})$. 
In the weak-coupling approximation employed here, only resonant processes are allowed.

The cumulant generating function of the model can be derived by 
following a rigorous procedure \cite{EspositoReview,HavaNJP}.
Here, we employ a classical, intuitive derivation of the FCS; it agrees with the rigorous method 
under the weak-coupling, Markovian and secular approximations \cite{HavaNJP}. 
We begin by defining $\mathcal{P}_t(j,nE_{2,1})$ 
as the probability that by the (long) time $t$,
the thermal energy $nE_{2,1}$ had been absorbed by the system  (transferred
from the cold bath), and the system occupies the state $j=1,2,3$. 
Note that, for simplicity, we count energy exchanged with the cold bath only.
Here, $n$  is an integer since  under the weak coupling approximation heat is transferred into and out of the system in 
discrete quanta, in resonance with system's level spacing.
We readily write down an equation of motion for $\mathcal{P}_t(j,nE_{2,1})$ 
\cite{EspositoReview,Renjie,Yelena10},
\begin{widetext}
\bea 
\dot{ \mathcal P}_t(1,nE_{2,1})
& = &
- \mathcal P_t(1,nE_{2,1}) (k_{1\rightarrow 2}^C +  k_{1\rightarrow 3}^H ) +
\mathcal P_t(2,(n+1)E_{2,1}) k_{2\rightarrow 1}^C +
\mathcal P_t(3,nE_{2,1}) k_{3\rightarrow 1}^H,
\nonumber\\
\dot{ \mathcal P}_t(2,nE_{2,1})
& = & - \mathcal P_t(2,nE_{2,1})
(k_{2\rightarrow 1}^C +  k_{2\rightarrow 3}^W ) +
\mathcal P_t(1,(n-1)E_{2,1}) k_{1\rightarrow 2}^C +
\mathcal P_t(3,nE_{2,1})k_{3\rightarrow 2}^W,
\nonumber\\
\dot{ \mathcal P}_t(3,nE_{2,1})
& = & - \mathcal P_t(3,nE_{2,1})
(k_{3\rightarrow 1}^H +  k_{3\rightarrow 2}^W ) +
\mathcal P_t(2,nE_{2,1}) k_{2\rightarrow 3}^W +  \mathcal P_t(1,nE_{2,1})
k_{1\rightarrow 3}^H.
\label{eq:Pwweak}
\eea
\end{widetext}
The equation is Fourier-transformed with the counting field
$\chi$, and we define the counting field dependent population 
\bea
|p(\chi,t)\rangle =
\begin{pmatrix}
\sum_{n=-\infty}^{\infty} \mathcal P_t(1,nE_{2,1})e^{inE_{2,1}\chi}  
\\ \sum_{n=-\infty}^{\infty} \mathcal
P_t(2,nE_{2,1})e^{i nE_{2,1}\chi}  
\\ \sum_{n=-\infty}^{\infty} \mathcal
P_t(3,nE_{2,1})e^{i nE_{2,1}\chi} 
\end{pmatrix}
\nonumber\\
\label{eq:zW}
 \eea
Note: the sign convention in the Fourier transform dictates the sign convention for the heat current, with
positive current corresponding to heat absorbed by the system. 
Using Eq. (\ref{eq:Pwweak}), we write down the set of differential equations,
\bea \dot p(\chi,t)=
 {\mathcal L}(\chi) p(\chi,t) 
\label{eq:ZW}
\eea
with the rate matrix (Liouvillian)
%
%
\bea
{\cal L}(\chi) = 
\begin{pmatrix} 
-k_{1 \to 2}^C - k_{1 \to 3}^H 					& k_{2 \to 1}^C e^{-i\chi E_{2,1}}  & k_{3 \to 1}^H\\
k_{1 \to 2}^C e^{i\chi E_{2,1}} 	& -k_{2 \to 1}^C - k_{2 \to 3}^W  & k_{3 \to 2}^W\\
k_{1 \to 3}^H 		&		k_{2 \to 3}^W 		& 	-k_{3 \to 1}^H - k_{3 \to 2}^W
\end{pmatrix}. \nonumber\\
\eea
The characteristic function is given by ${\mathcal Z}(\chi,t)=\langle I |p(\chi,t)\rangle$
with the identity vector $\langle I| $.
We now employ Eq. (\ref{eq:cooling}) to determine the cooling condition for the setup. 
We calculate the derivatives of the elements in ${\mathcal L} (\chi)$,
\bea
\frac{\partial {\cal L}(\chi)}{\partial (i \chi)} \Bigg|_{\chi=0} = 
\begin{pmatrix} 
0				& -E_{2,1}\, k_{2 \to 1}^C  & 0 \\
E_{2,1} \, k_{1 \to 2}^C  	& 0 & 0\\
0&0&0
\end{pmatrix}. 
\nonumber\\
\label{eq:derivA}
\eea
Again, we remark on the intuitive form of this matrix, pointing to the cooling processes.
It can be constructed without prior knowledge of the FCS method. 
The cooling condition is 
\bea
\trace \left[ 
\begin{pmatrix} 
C_{11} & C_{21} & C_{31} \\
C_{12} & C_{22} & C_{32} \\
C_{13} & C_{23} & C_{33}
\end{pmatrix}
\begin{pmatrix} 
0				& -E_{2,1}\, k_{2 \to 1}^C  & 0 \\
E_{2,1} \, k_{1 \to 2}^C  	& 0 & 0\\
0&0&0
\end{pmatrix}
 \right] >0, \nonumber
\eea
The matrix filled with the cofactors $C_{ij}$ is the adjugate matrix of the Liouvillian, ${\mathcal L}(0)$. 
This cooling condition condenses into
\bea
C_{21} e^{-\beta_C E_{2,1}} - C_{12} > 0.
\label{eq:condA}
\eea
If we interpret the cofactors as effective population,
 $C_{21}\leftrightarrow p_1$ and $C_{12}\leftrightarrow p_2$, we rationalize the cooling condition
as (nonequilibrium) deviations from the detailed balance relation.
We now evaluate the cofactors,
\bea
C_{21} &=& k^C_{2 \to 1} k^W_{3 \to 2} +  k^C_{2 \to 1} k^H_{3 \to 1} + k^H_{3 \to 1} k^W_{3 \to 2} e^{-\beta_W E_{3,2}} ,\nonumber \\
C_{12} &=& k^C_{2 \to 1} k^W_{3 \to 2} e^{-\beta_C E_{2,1}}+  k^C_{2 \to 1} k^H_{3 \to 1} e^{-\beta_C E_{2,1} } \nonumber \\&+& k^H_{3 \to 1} k^W_{3 \to 2} e^{-\beta_H E_{3,1}},
\eea
and substitute them in Eq. (\ref{eq:condA}). Altogether we get
\bea
 e^{-\beta_W E_{3,2}} e^{-\beta_C E_{2,1}} - e^{-\beta_H E_{3,1}} > 0. 
\label{eq:conditionideal}
\eea
This inequality can be simplified to arrive at the well-known cooling condition 
\bea
\frac{E_{2,1}}{E_{3,1}} < \frac{\beta_H-\beta_W}{\beta_C-\beta_W}. 
\label{eq:conditionidealfinal}
\eea
The formalism provides the cooling current, and
following similar steps, we get the heat absorbed from the work reservoir as well, 
thus the cooling coefficient of performance, $\eta\equiv \langle J_C\rangle/ \langle J_W \rangle = E_{2,1}/E_{3,2}$.
Combining this with Eq. (\ref{eq:conditionidealfinal}) we conclude that 
$\eta  \leq \eta_c$ with the Carnot cooling efficiency $\eta_c = (\beta_H-\beta_W)/(\beta_C-\beta_H)$.
Since one can tune parameters to reach the maximal cooling performance, 
we refer to this model as `ideal'.
This concludes the derivation of the cooling window for Model A presented in Fig. \ref{fig:contours}(a).

\vspace{3mm}


\subsection{Non-ideal QAR: Model B}

We analyze here Model B, as described and presented in Fig. \ref{fig:contours}(b). The Hamiltonian is given by
\bea
\hat{H} &=& \sum_{j=1}^3 E_j \ket{j}\bra{j} + \sum_{\substack {\mu,k}} \omega_{\mu,k} \, \hat a^{\dagger}_{\mu,k} \hat a_{\mu,k}   
\nonumber \\%
&+& \sum_{\mu\in C, W, H} \sum _{i > j}\hat B^\mu_{ij} (\ket{i}\bra{j} + \ket{j}\bra{i}).
\eea
In this model, each transition in the system is coupled to the three reservoirs, albeit we distinguish between
`dominant' couplings as in the ideal QAR model, and secondary (weaker) couplings.
Following the standard steps as performed for the ideal model,
we arrive at an  equation of motion for the counting field-dependent population,
$\dot p(\chi,t)=  {\mathcal L}(\chi) p(\chi,t)$ with the Liouvillian
\begin{widetext}
\bea
{\cal L}(\chi) =
\begin{pmatrix}
-\sum_{\mu=C,H,W} (k_{1 \to 2}^\mu+ k_{1 \to 3}^\mu)            & k_{2 \to 1}^C e^{-i\chi E_{2,1}}+k_{2 \to 1}^H+k_{2 \to 1}^W    &    k_{3 \to 1}^Ce^{-i\chi E_{3,1}}+k_{3 \to 1}^H+k_{3 \to 1}^W\\
k_{1 \to 2}^C e^{i\chi E_{2,1}}+k_{1 \to 2}^H+k_{1 \to 2}^W     & -\sum_{\mu=C,H,W} (k_{2 \to 1}^\mu+ k_{2 \to 3}^\mu)  & k_{3 \to 2}^Ce^{-i\chi E_{3,2}}+k_{3 \to 2}^H+k_{3 \to 2}^W\\
k_{1 \to 3}^C e^{i\chi E_{3,1}}+k_{1 \to 3}^H+k_{1 \to 3}^W             &               k_{2 \to 3}^C e^{i\chi E_{3,2}}+k_{2 \to 3}^H+k_{2 \to 3}^W             &       -\sum_{\mu=C,H,W} (k_{3 \to 1}^\mu+ k_{3 \to 2}^\mu)
\end{pmatrix}. \nonumber\\
\eea
\end{widetext}
Recall that counting is performed only at the cold bath.
The derivative is given as
\bea
\frac{\partial{\cal L}(\chi)}{\partial (i\chi)}\Bigg|_{\chi=0} =
\begin{pmatrix}
0               & -E_{2,1} k_{2 \to 1}^C     &  -E_{3,1} k_{3 \to 1}^C \\
E_{2,1} k_{1 \to 2}^C   & 0 & -E_{3,2} k_{3 \to 2}^C \\
E_{3,1} k_{1 \to 3}^C           &               E_{3,2} k_{2 \to 3}^C   &       0
\end{pmatrix} \nonumber\\
\eea
which generalizes (\ref{eq:derivA}), encompassing additional cold-bath mediated heat exchange processes.
Using Eq. (\ref{eq:current}), we obtain a condition on cooling,
\bea
&&E_{2,1}k^C_{2\to1}(C_{21}e^{-\beta_CE_{2,1}} - C_{12} ) \nonumber \\
&&+ E_{3,1}k^C_{3\to1}(C_{31}e^{-\beta_CE_{3,1}} - C_{13} ) \nonumber \\
&&+ E_{3,2}k^C_{3\to2}(C_{32}e^{-\beta_CE_{3,2}} - C_{23} )> 0.
\label{eq:coolingC}
\eea
%

The three contributions 
arise due to the coupling of the cold bath to the three transitions, $ |1\rangle  \leftrightarrow |2\rangle $,
$|1\rangle  \leftrightarrow |3\rangle $, and  $|2\rangle  \leftrightarrow |3\rangle $.
We begin by analyzing the second term in Eq. (\ref{eq:coolingC}), and we  write it down in terms of the rate constants,
%
\bea
&&C_{31} e^{-\beta_C E_{3,1}} - C_{13} =
\nonumber \\
&& \sum_{\mu,\nu} \Big[ k_{2 \to 1}^\mu k_{3\to2}^\nu (e^{-\beta_C E_{3,1}} - e^{-\beta_\mu E_{2,1}-\beta_\nu E_{3,2}}) \nonumber \\
&&+ k_{2 \to 1}^\mu k_{3\to1}^\nu (e^{-\beta_C E_{3,1}} - e^{-\beta_\nu E_{3,1}}) \nonumber \\
&&+ k_{3 \to 2}^\mu k_{3\to1}^\nu e^{-\beta_\mu E_{3,2}} (e^{-\beta_C E_{3,1} }- e^{-\beta_\nu E_{3,1}})  \Big]
\nonumber \\
\label{eq:C31} 
\eea
Since $\beta_C>\beta_H>\beta_W$, this contribution is always negative hindering cooling through the $1\leftrightarrow3$ transition.
Next, we consider the first row, 
$(C_{21}e^{-\beta_C E_{2,1} }- C_{12})$ in Eq. (\ref{eq:coolingC}) and
 write it explicitly in terms of the rate constants. We find that it includes heat leaks that
are always negative, as well as one term that could be positive---depending on the energy structure,
\bea
E_{2,1} k^C_{2\to1} k^W_{3\to2} k^H_{3\to1} (e^{-\beta_CE_{2,1}-\beta_WE_{3,2}}-e^{-\beta_H E_{3,1}}).
\nonumber\\
\label{eq:C1}
\eea
Similarly, the only potentially-positive combination
in the last row  of Eq. (\ref{eq:coolingC}),
$(C_{32}e^{-\beta_C E_{3,2}} - C_{23})$, is
\bea
E_{3,2} k^C_{3\to2} k^W_{2\to1} k^H_{3\to1} (e^{-\beta_WE_{2,1}-\beta_CE_{3,2}}-e^{-\beta_H E_{3,1}}).
\nonumber\\
\label{eq:C2}
\eea
Overall, we can write down the current extracted from the cold bath as
\bea
\langle J\rangle = \sum_{i>j}F_{cyc}^{i,j} + \sum_{\mu, i>j}F_{leak,\mu}^{i,j}.
\eea
with 
\bea
F_{leak,\mu}^{i,j}\propto e^{-\beta_C E_{i,j}}-  e^{-\beta_{\mu} E_{i,j}}
\eea
and
\bea
F_{cyc}^{2,1}&\propto& e^{-\beta_CE_{2,1}} e^{-\beta_WE_{3,2}}-e^{-\beta_HE_{3,1}}
\nonumber\\
F_{cyc}^{3,2}&\propto& e^{-\beta_CE_{3,2}} e^{-\beta_WE_{2,1}}-e^{-\beta_HE_{3,1}} 
\label{eq:condc12}
\eea
$F_{cyc}^{3,1}$ is given by the first contribution in Eq. (\ref{eq:C31}), and it is negative.

We conclude that there are two possible cooling cycles in the three level model: 
(i) $F_{cyc}^{2,1}$: The cold bath excites the system allowing the transition $\ket{1}\to\ket{2}$.  
The work bath pumps heat and induces the transition $\ket{2}\to\ket{3}$.
Heat is released to the $H$ bath with the system relaxing to $\ket{1}$.
(ii)  $F_{cyc}^{3,2}$: The cold bath excites the system, $\ket{2}\to\ket{3}$.  
The work bath induces the transition $\ket{1}\to\ket{2}$.
Heat is released to the $H$ bath with the system relaxing to $\ket{1}$.
These two circuits compete (``internal dissipation") and it is impossible to maximize the cooling performance simultaneously
and achieve maximal (Carnot) cooling:
Trying to optimize cycle (i) leads to suppression of the cooling current through cycle (ii), and vice versa 
\cite{Correa2015PRE,AlonsoNJP17}.
Particularly, conditions (\ref{eq:condc12}) are organized into the cooling conditions
(\ref{eq:condc1})-(\ref{eq:condc2}),  illustrating the conflicting requirements that the different cycles impose on the device.

\vspace{5mm}
\subsection{Non-ideal QAR: Models C and D}
We analyze here Models C and D, as described in Fig. \ref{fig:contours}(c)-(d).
Recalling, in Model C (D) the hot (work) bath leaks energy directly to the cold bath;
the cold bath couples at the $\ket 1 \to \ket 2$ transition. 
The model's Hamiltonian is given by
\bea
\hat{H} &=& \sum_{j=1}^3 E_j \ket{j}\bra{j} + \sum_{\substack {\mu,k}} \omega_{\mu,k} \, \hat a^{\dagger}_{\mu,k} \hat a_{\mu,k}   \nonumber \\%
&+& \hat B^C_{12} (\ket{1}\bra{2} + \ket{2}\bra{1})
+ \hat B^H_{13} (\ket{1}\bra{3} + \ket{3}\bra{1})  \nonumber \\ 
&+& \hat B^W_{23} (\ket{2}\bra{3} + \ket{3}\bra{2}) + \hat B^l_{12} (\ket{1}\bra{2} + \ket{2}\bra{1})
\nonumber\\
\eea
The last term describes the `leaky' ($l$) coupling: $l=H$ ($W$) in Model C (D).

Following similar definitions as for the ideal model, we arrive at the equation of motion for the
counting field-dependent populations,
$\dot p(\chi,t)=  {\mathcal L}(\chi) p(\chi,t)$ with the Liouvillian
\begin{widetext}
\bea
{\cal L}(\chi) = 
\begin{pmatrix} 
-k_{1 \to 2}^C - k_{1 \to 3}^H -k_{1 \to 2}^l		& k_{2 \to 1}^C e^{-i\chi E_{2,1}}+k_{2 \to 1}^l  & k_{3 \to 1}^H\\
k_{1 \to 2}^C e^{i\chi E_{2,1}}+k_{1 \to 2}^l 	& -k_{2 \to 1}^C - k_{2 \to 3}^W -k_{2 \to 1}^l  & k_{3 \to 2}^W\\
k_{1 \to 3}^H 		&		k_{2 \to 3}^W 		& 	-k_{3 \to 1}^H - k_{3 \to 2}^W
\end{pmatrix}. \nonumber\\
\eea
\end{widetext}
The derivative matrix $\partial {\cal L}/\partial (i\chi)|_{\chi=0}$ is the same as in the ideal case,
resulting in a formally identical cooling condition,
\bea
C_{21}e^{-\beta_CE_{2,1}} - C_{12} > 0,
\label{eq:coolingB}
\eea
yet with different cofactors $C_{21}$ and $C_{12}$,
\bea
C_{21} &=& k^C_{2 \to 1} k^W_{3 \to 2} +  k^C_{2 \to 1} k^H_{3 \to 1} + k^H_{3 \to 1} k^W_{3 \to 2} e^{-\beta_W E_{3,2}} ,\nonumber \\
&+& k^l_{2 \to 1} k^W_{3 \to 2} +  k^l_{2 \to 1} k^H_{3 \to 1}\nonumber \\
C_{12} &=& k^C_{2 \to 1} k^W_{3 \to 2} e^{-\beta_C E_{2,1}}+  k^C_{2 \to 1} k^H_{3 \to 1} e^{-\beta_C E_{2,1} } \nonumber \\&+& k^H_{3 \to 1} k^W_{3 \to 2} e^{-\beta_H E_{3,1}}+ k^l_{2 \to 1} k^W_{3 \to 2} e^{-\beta_l E_{2,1}}\nonumber \\ &+&  k^l_{2 \to 1} k^H_{3 \to 1} e^{-\beta_l E_{2,1} }. \nonumber 
\eea
We plug these expressions back into the cooling condition (\ref{eq:coolingB})
and get
\bea
&&\left[ e^{-\beta_C E_{2,1}-\beta_W E_{3,2}}-e^{-\beta_H E_{3,1}} \right] 
\nonumber \\
&&+ \left[k_{2 \to 1}^l \left( \frac{1}{k^H_{3\to1}} + \frac{1}{k^W_{3\to2}} \right) 
( e^{-\beta_C E_{2,1}} - e^{-\beta_l E_{2,1}}  ) \right]>0
\nonumber\\
\eea
The first square-bracket term corresponds to the cooling condition of the ideal model A.
The second square bracket describes the heat flow from a hot bath ($H$ or $W$)  to the cold bath. 
Since the second term is always negative, we conclude that (as expected)
the cooling condition is more limited in Model C or D, compared to the ideal design, Model A.




\begin{thebibliography}{0}
\bibitem{commentJ}
In the standard approach, the heat current from  the $\mu$ bath is identified from the change of  the  energy of the system, 
$d\langle \hat H_S\rangle/dt$, which leads to $\langle J_C\rangle = \sum_j  E_j ({\mathcal L}_{C}p^{ss})_j$
with $p^{ss}$ as the vector of population in steady state, obtained by solving the MQME in
the long time limit.
\bibitem{commentN3}
For a three level system, $a_{1}=-{\rm Tr}[{\cal L}]$, which does not depend on $\chi$.
As for, $a_2$, if the cold bath (whose energy exchange we count) is separately coupled to a selected transition,
it includes multiplications of the form ${\cal L}_{i,j}(\chi) {\cal L}_{j,i}(\chi)$, with
${\cal L}_{i,j}\propto e^{i\chi\epsilon_{i,j}}$;  $\epsilon_{i,j}$ the transition energy.
\end{thebibliography}
\end{document}